\ifdefined\XeTeXversion
  \PassOptionsToPackage{xetex}{hyperref}
\else
  \pdfoutput=1
\fi
\documentclass[acmsmall]{acmart} 
\renewcommand\footnotetextcopyrightpermission[1]{} 
\AtBeginDocument{%
  }
    
\usepackage{algorithm}
\usepackage{algorithmic}

\usepackage{tabularx} 
\usepackage{array}

\usepackage{multirow} 
\usepackage{booktabs} 
\usepackage{makecell} 
\usepackage{array} 

\usepackage{ragged2e} 

\usepackage{caption}
\usepackage{array} 
\usepackage{cellspace} 
\usepackage{array} 
\usepackage{cellspace} 
\usepackage{colortbl} 
\definecolor{lightgray}{gray}{0.8} 

\usepackage[utf8]{inputenc}
\usepackage{listings}
\usepackage{xcolor}

\lstdefinelanguage{Solidity}{
    keywords={contract, function, public, payable, mapping, address, uint256, msg, sender}, 
    keywordstyle=\color{cyan!80!black}\bfseries, 
    morestring=[b]", 
    stringstyle=\color{orange!80!black}, 
    morecomment=[l]{//}, 
    morecomment=[s]{/*}{*/}, 
    commentstyle=\color{gray!60}\itshape, 
    sensitive=true 
}

\lstdefinelanguage{Rust}{
    keywords={fn,let,mut,contract, function, public, payable, mapping, address, uint256, msg, sender}, 
    keywordstyle=\color{cyan!80!black}\bfseries, 
    morestring=[b]", 
    stringstyle=\color{orange!80!black}, 
    morecomment=[l]{//}, 
    morecomment=[s]{/*}{*/}, 
    commentstyle=\color{gray!60}\itshape, 
    sensitive=true 
}

\begin{document}

\title{Exploring Vulnerabilities and Concerns in Solana Smart Contracts}

\footnotetext{These authors contributed equally to this work.}

\author{Xiangfan Wu}
\authornotemark[1] 
\affiliation{%
  \institution{Hainan University}
  \city{Haikou}
  \country{China}}

\author{Ju Xing}
\authornote{These authors contributed equally to this work.}
\affiliation{%
  \institution{Hainan University}
  \city{Haikou}
  \country{China}}
\email{yeshipeng35@gmail.com}

\author{Xiaoqi Li}
\affiliation{%
  \institution{Hainan University}
  \city{Haikou}
  \country{China}}
\email{csxqli@ieee.org}

\begin{abstract}
The Solana blockchain was created by Anatoly Yakovenko of Solana Labs and was introduced in 2017, employing a novel transaction verification method. However, at the same time, the innovation process introduced some new security issues. The frequent security incidents in smart contracts have not only caused enormous economic losses, but also undermined the credit system based on the blockchain. The security and reliability of smart contracts have become a new focus of research both domestically and abroad. This paper studies the current status of security analysis of Solana by researching Solana smart contract security analysis tools.
This paper systematically sorts out the vulnerabilities existing in Solana smart contracts and gives examples of some vulnerabilities, summarizes the principles of security analysis tools, and comprehensively summarizes and details the security analysis tools in Solana smart contracts. The data of Solana smart contract security analysis tools are collected and compared with Ethereum, and the differences are analyzed and some tools are selected for practical testing.
\end{abstract}

\maketitle

\pagestyle{plain} 

\section{Introduction}

Blockchain is a decentralized network that supports the composition of distributed records stored in immutable blocks into a continuously growing chain. Over the past decade, blockchain technology has evolved from the ledgers of cryptocurrencies such as Bitcoin and Monero to distributed computing platforms like Ethereum and EOS, which allow the deployment and execution of smart contracts. Smart contracts are decentralized programs deployed on the blockchain that can enforce agreements and protocols without involving any third party or establishing mutual trust. They provide a set of functions that can be invoked through transactions and executed by the blockchain's virtual machine (VM). Most smart contracts are written in high-level specialized programming languages such as Solidity, JavaScript, or Vyper and compiled into blockchain VM bytecode. For example, the Ethereum Virtual Machine (EVM) is the blockchain VM that executes smart contracts on the Ethereum platform. An important feature of smart contracts is their ability to perform financial operations using cryptocurrencies and valuable custom tokens such as ERC20 and ERC721. These applications expose assets to threats at both the contract and blockchain-system levels.\cite{BlockchainSecuritySurvey}

Due to the storage and transfer of valuable assets through smart contracts, they have become a priority target for attackers. Incidents involving the DAO and Parity wallets illustrate how implementation flaws can lead to asset theft or inaccessible funds. Existing surveys document the corresponding vulnerabilities, attacks, and defensive mechanisms.\cite{sok,SurveyEthSys} These incidents motivate security analysis adapted to each platform\textquotesingle s execution and state model.

Solana was introduced by Anatoly Yakovenko in 2017 as a blockchain architecture designed for high throughput. Its Proof of History (PoH) mechanism provides a verifiable sequence of events that assists transaction ordering and coordination. PoH should be distinguished from the consensus protocol that uses this ordering information.\cite{SolanaWhitepaper} The security implications of Solana programs therefore need to be considered in the context of its execution model rather than inferred directly from Ethereum defect classifications.

Solana programs are compiled through LLVM to Solana Bytecode Format (sBPF) and executed by the Solana virtual machine. Executable code and mutable state are separated: programs operate on data accounts supplied through instructions.\cite{SolanaPrograms} Rust is a principal language for developing these programs. Its ownership and type systems provide memory-safety guarantees for safe code, but do not establish the correctness of application-level authorization or financial logic.\cite{RustThe,RustBelt}

We have compiled a list of the major attacks on Solana blockchain smart contracts since February 2022, as shown in Tab.~\ref{tab:Statistics on major attacks against Solana smart contracts}. For example, the Wormhole incident involved a forged sysvar account that bypassed signature verification, illustrating the importance of validating security-critical accounts.\cite{WormholeIncident} 
\begin{table}
  \caption{Statistics on major attacks against Solana smart contracts}
  \begin{tabular}{ccccc}
    \toprule
    no & Attack Time & Attack target & Loss Amount & Attack Method\\
    \midrule
    1 & 11/02/2022 & Solend & \$1,260,000 & Oracle Attack\\
    2 & 10/11/2022 & Mango & \$100,000,000 & Flash Loan \\
    3 & 10/12/2022 & TulipProtocol & \$2,500,000 & Mango Attack \\
    4 & 10/12/2022 & UXD Protocol & \$20,000,000 & Mango Attack \\
    5 & 08/29/2022 & OptiFi & 661,000 USDC & Operational Error \\
    6 & 07/28/2022 & Nirvana & \$3,500,000 & Flash Loan \\
    7 & 07/03/2022 & Crema Finance & \$1,682,000 & Flash Loan \\
    8 & 03/30/2022 & Jet Protocol & Unknown & Unknown \\
    9 & 03/23/2022 & Cashio & \$52,027,994 & Hacker bypassed unverified accounts \\
    10 & 02/02/2022 & Wormhole & 120,000 ETH & Developer enabled forged signatures via deprecated function. \\

  \bottomrule
\end{tabular}

  \label{tab:Statistics on major attacks against Solana smart contracts}
  
\end{table}

\subsection{Related Research}
Some previous literature has been published to investigate the security of smart contracts, but they have different perspectives from this survey.Atzei et al. \cite{sok} presented the first systematic exposition of Ethereum security vulnerabilities, classifying them into three levels: Solidity layer, EVM bytecode layer and blockchain layer, and demonstrated six influential attacks in different application scenarios. In contrast, this paper mainly focuses on the security analysis and defense methods of Solana blockchain smart contracts rather than the classification of program vulnerabilities. Jiachi et al. \cite{DefiningSmart} conducted an empirical investigation, systematically studying the defects of smart contracts on the Ethereum platform from five aspects: security, usability, performance, maintainability, and reusability. They collected and analyzed posts related to smart contracts on Ethereum.StackExchange and smart contracts with actual problems, defining 20 types of contract defects and 5 related impacts. Zou et al.\cite{SCDevelop} carried out an exploratory study, which illustrates the current state and the potential challenges of smart contract development. Specifically, they conducted semi-structured interviews with 20 developers and professionals, followed by a survey of 232 practitioners to confirm the five conclusions from these interviews, with a focus on smart contract development. Furthermore, Zhang et al. \cite{FramworkAnd} proposed a new framework for classifying vulnerabilities of smart contracts and constructed a dataset of 176 defective smart contracts.  Vacca et al. \cite{SystemLitera}conducted a systematic review of technologies and tools to address software engineering challenges in blockchain-based applications by analyzing 96 articles and articles. Previous surveys summarize the security and development issues of smart contracts, while this article focuses on solutions for the security analysis and defense of Solana blockchain smart contracts.

Some surveys have considered solutions for the analysis and defense of smart contract security. Huashan et al. \cite{SurveyEthSys}presented a comprehensive and detailed investigation of the security of the Ethereum system, including vulnerabilities, attacks, and defense measures. They discussed 44 types of vulnerability and described the history, causes, strategies, and direct impacts of 26 types of attack. Regarding defense measures, they listed 47 defense mechanisms and provided best practices to guide contract development. Wang and He et al. \cite{SecurityEnhance} reviewed six vulnerability detection methods and privacy protection technologies on three platforms (i.e., Ethereum, Hyperledger Fabric and Corda) and summarized the several commonly used tools employed by each method. Di Angelo et al. \cite{SurveyTool}investigated the availability, maturity level, adopted methods, and detection of security issues of 27 Ethereum smart contract analysis tools. They examined the availability and functionality of these tools and compared their characteristics in a structured way.

Li et al. survey threats, real attacks, and security enhancements across blockchain systems.\cite{BlockchainSecuritySurvey} This broader perspective complements the Ethereum-focused literature, while account validation and the Rust-based execution environment require platform-specific treatment for Solana.

The approach in this paper will focus more on the security analysis and defense methods of Solana blockchain smart contracts. This paper will investigate the availability, maturity level, adopted methods, and detection of security issues of multiple smart contract security analysis tools specifically designed for the Solana blockchain.




\section{Solana Ecosytem}

In this chapter, we have collected relevant data, and conducted detailed data analysis. This will better help us understand the differences and current status between the Solana ecosystem and the Ethereum ecosystem.

\subsection{Data Collection}
This paper analyzes the activity of blockchain platforms by comparing indicators such as the number of projects, issues, and stars of different smart contract projects on GitHub. This method can reflect the performance of different blockchain platforms in community building, developer support, and ecosystem construction. Stars and issues provide descriptive indicators of repository visibility and activity, but neither directly measures detection accuracy or security quality. Therefore, this data-driven statistical analysis method can provide valuable information to help people better understand and evaluate the environments of smart contract security analysis tool platforms on different blockchain platforms, as shown in Tab. ~\ref{tab:Comparison of Solana and Ethereum Programs Based on Issues and Stars}

\begin{table}
  \caption{Comparison of Solana and Ethereum Programs Based on Issues and Stars}
  \begin{tabular}{ccccccc}
    \toprule
    
    No. & Solana Program & issue & star & Ethereum Program & issue & star\\
    
    \midrule
    
    1 & trdelnik & 2 & 77 & Oyente & 63 & 1229 \\
    2 & anchor-UI & 0 & 26 & Mythril & 92 & 3132\\
    3 & Blockworks Checked Math & 0 & 6 & Securify & 32 & 489\\
    4 & cargo-geiger & 26 & 1200 & Manticore & 237 & 3406 \\
    5 & solana-poc-framework & 4 & 171 & Slither & 411 & 4072 \\
    6 & sol-ctf-framework & 1 & 43 & Maian & 27 & 524 \\
    7 & vipers & 0 & 129 & Vandal & 28 & 137 \\
    
  \bottomrule
\end{tabular}
  \label{tab:Comparison of Solana and Ethereum Programs Based on Issues and Stars}
\end{table}

Tool availability should also be distinguished from measured detection effectiveness. For example, vipers provides reusable validation helpers for Solana programs, while Trident provides a framework for fuzzing Solana programs.\cite{vipers,Trident} Their official repositories document their purpose and usage; these sources do not establish the historical repository counts in our dataset.

In actual practice, using code volume to measure the size of a project is no longer very accurate, because projects often contain a large amount of third-party library code and references, or to make the project structure clearer, the core is stripped out of the project. Therefore, we use issues and stars to measure the activity of projects.

\begin{figure}[h]
    \centering
    \includegraphics[width=0.8\textwidth]{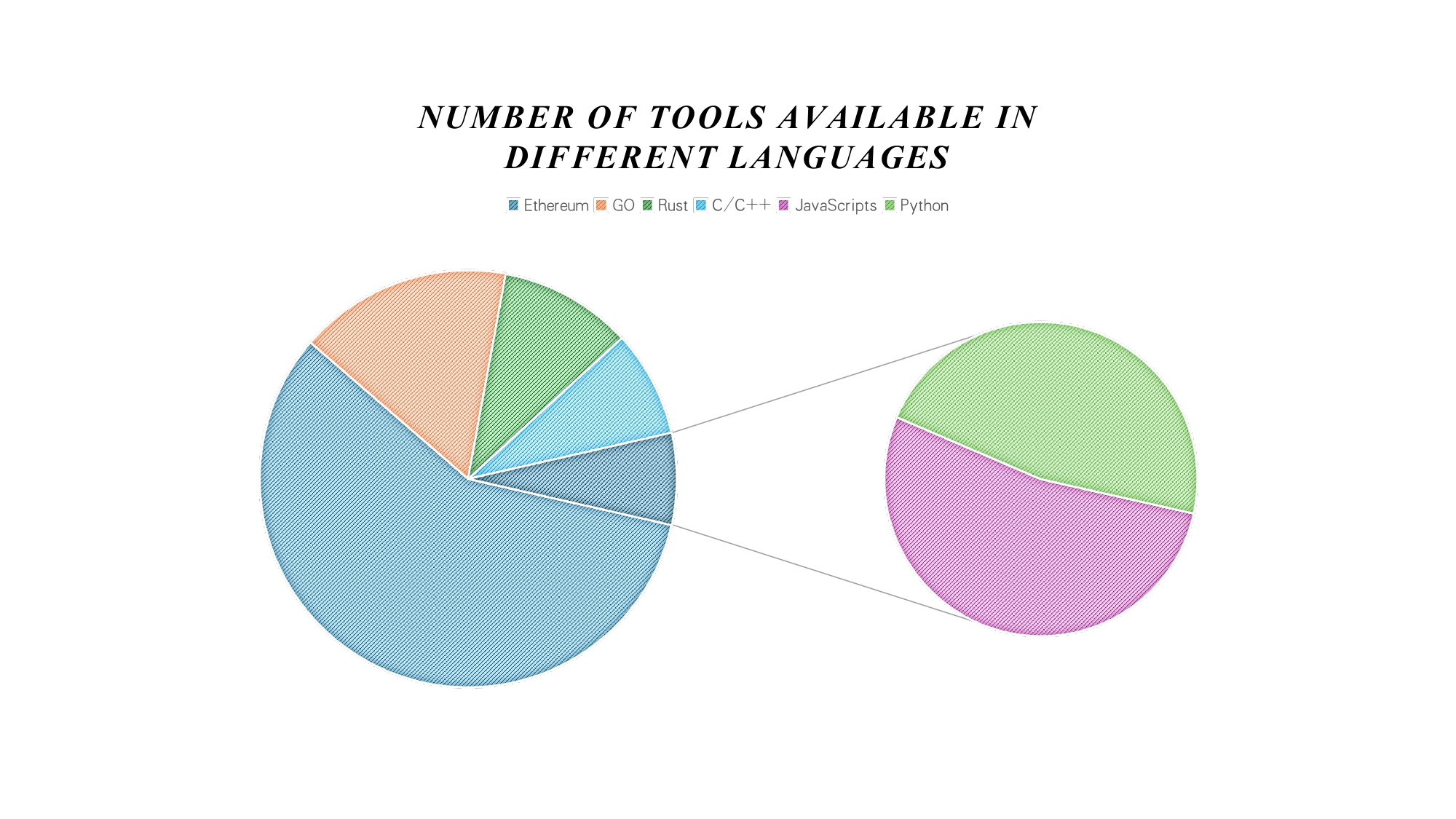}
    \caption{ Number of Tools Available in Different Languages}
    \label{fig:Number of Tools Available in Different Languages}
\end{figure}

As shown in Fig.~\ref{fig:Number of Tools Available in Different Languages}, our collected inventory contains 113 tools supporting Ethereum smart contracts, 12 supporting Solana smart contracts, and 13 supporting multiple platforms or general analysis tasks. These figures describe the collected sample rather than an exhaustive, continuously updated census of available tools. Comparisons should account for the inclusion criteria and collection period.

Ethereum can become the mainstream platform for blockchain smart contract security analysis and development for the following reasons: Firstly, Ethereum was launched relatively early, which has attracted more attention and support. Secondly, Ethereum has a large scale and a huge user base, which has increased its influence. Thirdly, the Ethereum ecosystem is relatively complete, providing more support and resources for application developers. Finally, at the programming language level, Ethereum uses the Solidity language. Many other blockchain platforms also adopt similar virtual machines and programming languages, which means that developers can more easily migrate their applications to other platforms.

Among the 12 Solana security analysis tools, 7 are open-source and the remaining 5 are closed-source.

In terms of type comparison, we classify tools into static analysis, dynamic analysis, and symbolic analysis, etc. 

For static analysis tools, they can be divided into two types according to their characteristics: specialized type detection tools and comprehensive detection tools. Specialized type detection tools refer to those that are designed to detect only a specific domain or technology. For example, Blockworks Checked Math and cargo-audit mentioned above. Blockworks Checked Math focuses on checking mathematical operations, while cargo-audit focuses on checking unsafe library references in Rust. These tools are often optimized and designed for specific domains, so they have a high detection accuracy for problems in the corresponding domains. Comprehensive detection tools, on the other hand, refer to those that can scan and detect all the code in an entire project or codebase, such as Kudelski Semgrep. These tools can perform global detection and analysis and provide a wider detection range. However, due to their broad detection scope, they may not be as accurate as specialized type detection tools in detecting problems within specific domains. Therefore, in practical applications, it is necessary to choose the appropriate static analysis tool according to actual needs. If developers want to detect a specific domain or technology, they can choose specialized type detection tools; if they need to perform comprehensive static analysis of the code, they can choose comprehensive detection tools.

\subsection{Data Analysis}

From the above data analysis and comparison, it can be found that smart contract security analysis tools in the Ethereum environment have higher popularity and more issues. This is mainly because Ethereum, as one of the earliest smart contract platforms, has a wider range of users and application scenarios, attracting more attention and followers.

In addition, the types of smart contract security analysis tools in Ethereum are more diverse. In addition to common static analysis tools and dynamic analysis tools, Ethereum also includes binary analysis tools, deep learning analysis tools, and other types of security analysis tools. These tools can conduct more in-depth and comprehensive security detection and analysis of contracts, discovering more potential vulnerabilities and security risks. In contrast, the variety of Solana smart contract security analysis tools is relatively small and still needs further improvement and development. However, since Solana can leverage the advantages of the Rust language ecosystem, it also has some relatively popular tools.

It was also found that there are fewer large open-source analysis tools in the Solana ecosystem. For example, SEC (i.e., Soteria) is the only symbolic analysis tool. However, with its commercialization, it is currently not open-source and does not provide a free version.

\section{Principles of Vulnerability}

Smart contract vulnerabilities often arise for a variety of reasons, including bad practices, coding errors, and more. These vulnerabilities can lead to smart contracts executing unintended behaviors, such as transferring funds to unauthorized accounts, replay attacks, data tampering, and even denial-of-service attacks. Furthermore, vulnerabilities in smart contracts can be exploited to attack the entire blockchain network, thereby affecting the whole network. \cite{DefiningSmart} Therefore, the security of smart contracts is crucial. To prevent vulnerabilities, developers should follow best practices and avoid introducing vulnerabilities as much as possible after development is completed, such as conducting thorough testing and audits during the development process, using reliable development tools, and limiting access and execution permissions, etc.
This chapter will delve into the various types and principles of smart contract vulnerabilities and how to prevent and fix these vulnerabilities, in order to help readers better understand and protect the security of smart contracts.

\subsection{Lack of Check}

The underlying principles of Solana are different from those of Solidity. It uses a different language—Rust (as opposed to Solidity in Ethereum)—and decouples code and data. It defines all addresses as accounts, and each account will store either funds or programs. Programs allow anyone to call them by simply providing input parameters, but this can also lead to serious consequences such as theft of funds if proper identity checks are forgotten.

\subsubsection{Lack of Signer Check}

The function described in Listing 1 is designed to update the administrator of the contract. Its primary purpose is to maintain the security and integrity of the contract by allowing only the current administrator to initiate changes to the administrator account. To achieve this, the function includes a comparison mechanism that checks whether the incoming account matches the current administrator account. However, this approach has a significant flaw: it lacks verification of whether the current administrator actually signed the operation.

The missing signature check is a documented Solana vulnerability pattern.\cite{SolanaSecurityPitfalls} Anchor provides a signer constraint to express this validation.\cite{AnchorConstraints}

This omission creates a critical vulnerability. An attacker could exploit this weakness by passing the current administrator account as a parameter when calling the function, while simultaneously setting their own account as the new administrator. Since the function does not verify whether the current administrator's signature is present or valid, it cannot confirm that the operation was indeed authorized by the legitimate administrator. As a result, the attacker can successfully replace the current administrator with a malicious account. This would grant the attacker full control over the contract, allowing them to manipulate its functionality, steal assets, or disrupt its operations.

\begin{lstlisting}[language=Rust,caption={Example of Lacking Singer Check}]
fn update_admin(program_id: &Pubkey, accounts: &[AccountInfo]) -> ProgramResult {
    let account_iter = &mut accounts.iter();
    let config = ConfigAccount::unpack(next_account_info(account_iter)?);
    let admin = next_account_info(account_iter)?;
    let new_admin = next_account_info(account_iter)?;
    // ...
    // Validate the config account...
    // ...
    if admin.pubkey() != config.admin {
        return Err(ProgramError::InvalidAdminAccount);
    }
    
    config.admin = new_admin.pubkey();
    
    Ok(());
}
\end{lstlisting}

\subsubsection{Lack of Ownership Check}

An account\textquotesingle s owner field identifies the program permitted to modify its data; it is distinct from the user or authority stored inside that data.\cite{SolanaAccounts} Before trusting a configuration account, a program must verify its expected owner and separately validate the authority used by the instruction. Without the owner check, an attacker can supply a configuration account containing a forged administrator field.\cite{SolanaSecurityPitfalls} Anchor exposes owner and authority constraints for these checks.\cite{AnchorConstraints}

\subsubsection{Missing rent-exemption check}

Account creation must provide the minimum lamport balance required for the allocated data size. Solana\textquotesingle s account documentation describes this storage balance, and Anchor initialization constraints enforce rent exemption unless explicitly skipped.\cite{SolanaAccounts,AnchorConstraints} A program should therefore validate the intended account size and funding requirements when creating or initializing accounts. The relevant protocol version determines account-loading and storage-balance behavior.

\subsection{Conflation}

\subsubsection{Solana Account Confusion}

Typically, contracts require multiple types of account to store state and data. Each account type serves a different role and function, ensuring proper storage and access to data. Simply checking the account owner is not sufficient, as it does not fully guarantee that the correct type of account is being operated on. Account discriminators and constraints on addresses or PDA seeds help distinguish the intended account type and instance.\cite{AnchorConstraints} If the account type is incorrect, the contract's functionality may exhibit unintended behavior or security issues. Therefore, each account passed into the contract must be validated to ensure it is indeed of the expected type.

Additionally, when the contract is updated and the data format of the account types is modified, extra care must be taken. Such changes may cause old data to be unreadable or improperly processed, so verifying the account's data format version becomes crucial. For example, if the data structure changes, a new data type or version number can be introduced for each modified account type to differentiate it. This ensures that the updated contract can correctly read and process the data in the accounts, avoiding errors or vulnerabilities due to format mismatches.

\subsubsection{Cross-Instance Confusion Re-initialization Attack}

Re-initialization can reset an existing account\textquotesingle s authority or application state when a program fails to distinguish initialization from subsequent use. It is not sufficient to clear state indiscriminately: initialization should be authorized, account addresses and seeds should identify the intended instance, and previously initialized accounts should not be reset through an unintended path. Anchor\textquotesingle s documentation explicitly warns that \texttt{init\_if\_needed} requires protection against re-initialization attacks.\cite{AnchorConstraints}

\subsection{Calculation Errors}

\subsubsection{Arithmetic Overflow/Underflow}

Rust is a systems programming language\cite{RustThe} that provides memory safety, concurrency, and high performance. However, integer overflow vulnerabilities are a potential issue in Rust. Integer overflow occurs when a computer performs an operation on an integer value and the result exceeds the range that the type can represent.For example, if a u8 type is used to represent an unsigned integer in the range 0-255, and an attempt is made to add 256 to a u8 variable, an integer overflow will occur, resulting in the value of the variable being 0. In Rust, when integer overflow is not properly handled, it can introduce serious security problems. For instance, malicious attackers might exploit integer overflow to bypass security checks or perform other harmful operations. Explicit checked arithmetic and checked conversions can prevent unintended wrapping in security-critical calculations.\cite{SolanaSecurityPitfalls}

Integer overflow vulnerabilities are a common type of vulnerability. For example, the \textit{BeautyChain} team announced on April 22, 2018, that the BEC token experienced abnormal fluctuations on that day. Attackers successfully obtained $10^{58}$ BECs by exploiting the vulnerability caused by integer overflow. In the attack event of this contract, the attacker executed the function \textit{batchTransfer}, which had an integer overflow vulnerability, to conduct transactions. \cite{DefiningSmart}

\begin{figure}[h]
    \centering
    \includegraphics[width=0.8\textwidth]{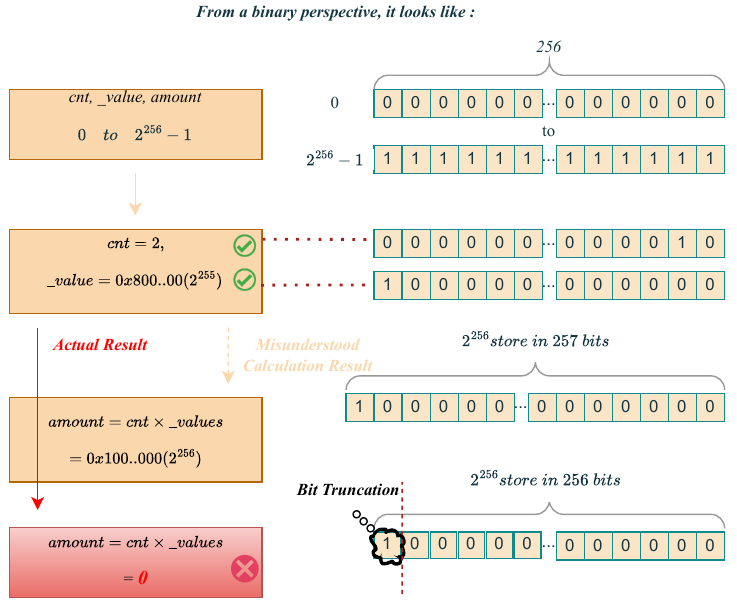}
    \caption{Interger Overflow in Binary Perspective }
    \label{fig:Interger Overflow in Binary Perspective}
\end{figure}

The \textit{batchTransfer} function Shown in Fig.\ref{fig:Interger Overflow in Binary Perspective} is used to transfer a certain amount of tokens \textit{\_value} to multiple addresses \textit{\_receivers}. The basic flow of the function is as follows: first, it calculates the total amount to be transferred \textit{amount}, then checks the transfer conditions, including the number of receivers \textit{cnt}, the transfer amount for each receiver \textit{\_value}, and whether the sender's balance is sufficient. If all conditions are met, the transfer is executed by deducting the specified amount from the sender's balance and distributing it to each receiver.

The root cause of the vulnerability lies in the multiplication operation, which could lead to an overflow. In Solidity, the maximum value for the \textit{uint256} type is $2^{256} - 1$. Assuming that \textit{cnt} has a maximum value of 20 (limited by require(cnt <= 20)), \textit{\_value} is not constrained. Therefore, an attacker can set \textit{\_value} to a very large number. For example, \textit{\_value} could be set to \textit{0x8000000000000000000000000000000000000000000000000000000000000000}, which is $2^{255}$.
\textit{}
When the number of receivers \textit{cnt = 2}, \textit{amount} will be calculated as \textit{$2 * 2^{255} = 2^{256}$}, which exceeds the \textit{uint256} representation range, causing an overflow. Due to the overflow, \textit{amount} will become 0. At this point, the balance check \textit{require(balances[msg.sender] >= amount)} will always pass because 0 is less than any balance. As a result, the balance check condition will not prevent the subsequent transfer, allowing the attacker to proceed with the transfer.

The critical impact of this vulnerability is that the attacker can bypass the balance check, successfully transferring assets to the receiver's account without actually deducting the amount from the sender's account. This allows the attacker to make illegal token transfers without having sufficient balance, leading to financial loss.

The following listing.2 is the specific implementation of that function.

\begin{lstlisting}[language=solidity,caption=Example of Interger Overflow]
function batchTransfer(address[] _receivers, uint256 _value) public whenNotPausedreturns (bool) {
  uint cnt = _receivers.length;
  uint256 amount = uint256(cnt) * _value;

  require(cnt > 0 && cnt <= 20);
  require(_value > 0 && balances[msg.sender] >= amount);

  balances[msg.sender] = balances[msg.sender].sub(amount);
  for (uint i = 0; i < cnt; i++) {
      balances[_receivers[i]] = balances[_receivers[i]].add(_value);
      Transfer(msg.sender, _receivers[i], _value);
  }
  return true;
}
\end{lstlisting}

\subsubsection{Numerical Precision Errors}

Financial calculations require explicit numeric representations and rounding rules. Integer division and fixed-point arithmetic can lose precision even when integer overflow is prevented. For example, Uniswap v2 represents oracle prices with a binary fixed-point format.\cite{UNISWAP} Security reviews should examine scaling factors, intermediate products, rounding direction, and accumulated error in the implemented arithmetic; these issues cannot be resolved solely by replacing calculations with general numerical algorithms.

\subsection{Unsafe Code}

\subsubsection{Using unsafe Rust Code}

Unsafe Rust permits operations such as raw-pointer dereferencing and calling unsafe functions, for which the programmer must establish the required safety conditions. It does not disable all compiler checks.\cite{UnsafeRust} RustBelt formalizes how verified unsafe components can expose safe interfaces, illustrating why the safety of abstractions depends on their implementation.\cite{RustBelt} In Solana programs, unsafe blocks and their dependencies should be reviewed for memory-safety invariants in addition to account authorization and business-logic requirements.

\subsubsection{Outdated Dependencies}

The Rust language and the Cargo package manager indeed simplify dependency management. However, dependencies may become outdated or contain known security vulnerabilities, which may affect the performance, stability, and security of the code. Therefore, it is necessary to update dependencies in a timely manner .
RustSec maintains advisories for Rust packages, and \texttt{cargo-audit} checks dependencies against this database.\cite{RustSec} Such checks help identify known vulnerable dependencies but do not establish the absence of unknown vulnerabilities.

\subsection{Logic Vulnerabilities}

Logic vulnerabilities in smart contracts refer to high-level semantic errors in the code of smart contracts, which may cause the contract to produce unpredictable results during execution or be exploited maliciously. These vulnerabilities may include problems with oracles and manipulation of transaction order. To ensure the reliability and security of smart contracts during execution, comprehensive logical analysis and testing of the contract are required to ensure its logical correctness and security. In this chapter, we will explore the possible logic vulnerabilities in smart contracts and introduce how to effectively identify and solve these problems.

Logical defects can also involve incorrect, incomplete, or unauthorized state transitions across interacting contracts. StateGuard constructs graph representations from smart-contract abstract syntax tree (AST) features and uses a graph convolutional network to detect state derailment defects in decentralized exchange (DEX) contracts.\cite{StateGuard} This work motivates state-sensitive analysis, although its reported evaluation should not be treated as a validation on Solana programs.

\subsubsection{Sandwich Attack}

A sandwich attack is a popular front-running technique in DeFi\cite{EliminatingSandwich}. The attacker finds a pending victim transaction and attempts to sandwich the victim between two transactions, forming a "sandwich" - style transaction. This strategy originates from the method of manipulating asset prices by buying and selling assets. The transparency of the blockchain and the delay in executing orders (usually in the case of network congestion) make front-running easier and significantly reduce the security of transactions. On Ethereum, visible pending transactions and competition over transaction ordering can enable front-running and miner extractable value (MEV).\cite{FlashBoys2} This observation should not be generalized into a claim that every blockchain exposes the same public mempool. Predatory traders will notice when a pending asset X transaction of a potential victim is used for asset Y, and they will buy asset Y before the victim. They know that the victim's transaction will increase the asset price, so they plan to buy asset Y at a lower price, let the victim buy it at a higher price, and finally sell the asset at a higher price.
Uniswap\textquotesingle s automated market maker illustrates the constant-product trading mechanism used by many decentralized exchanges.\cite{UNISWAP} The constant product market maker model is a liquidity provision method for decentralized exchanges. It keeps the product of a token pair constant and determines the price of each token based on supply and demand. When a user buys X tokens at P3 (i.e., sells Y tokens), the price of X tokens will rise and the price of Y tokens will fall.

\begin{figure}[h]
    \centering
    \includegraphics[scale=0.05]{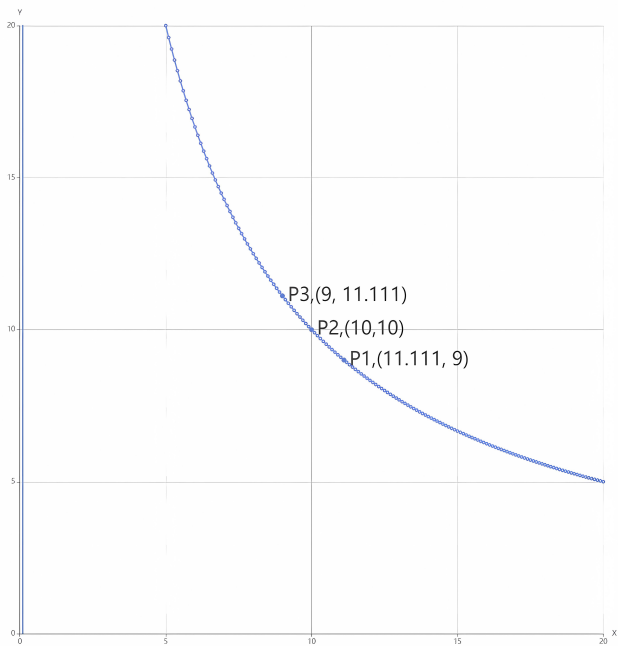}
    \caption{AMM Model While Z=100 }
    \label{fig:AMM}
\end{figure}

Fig .\ref{fig:AMM} illustrates a sandwich attack. When the user purchases X tokens, the attacker buys X tokens first, causing the price to slide from P3 to P2. When the user successfully purchases X tokens, the price slides back from P2 to P1. After the price reaches P1, the attacker sells the tokens purchased at P3. As a result, the attacker buys X tokens at the X coordinate of P3 and sells them at the X coordinate of P1, earning a profit from the price difference."

A sandwich attack means that when a user buys X tokens, the attacker front-runs and buys X tokens first, causing the price to slide from P3 to P2. When the user successfully buys X tokens, the price slides again from P2 to P1. After the price reaches P1, the attacker sells the tokens bought at P3 again. Thus, the attacker buys X tokens at the X-coordinate of P3 and sells X tokens at the X-coordinate of P1, obtaining the price spread.

Detection can complement preventive controls. GasTrace combines gas-related transaction features with a cascade classifier and graph attention network to identify sandwich-attack accounts on Ethereum.\cite{GasTrace} Applying this idea to Solana would require adapting its features and evaluating the model against Solana\textquotesingle s transaction and fee mechanisms.

\subsubsection{Oracle Attacks}

Oracles act as interfaces that connect Dapps with the external real world. They can call various external data resources, such as market prices, to provide the required data for Dapps. However, since oracles are not under the control of Dapps, there is a risk of being manipulated by hackers. Specifically, attackers may tamper with the returned results by manipulating external data sources or attacking the oracle, causing parameters such as prices in Dapps to deviate from the real world. If attackers can successfully exploit these vulnerabilities for arbitrage, they may obtain huge profits. Therefore, oracle attacks have become a major security hazard in the DeFi field. Flash loans can provide temporary capital for price manipulation, but flash loans and oracle attacks are not equivalent categories. A vulnerable protocol may use a manipulable price when valuing collateral or executing a trade, allowing an attacker to exploit the resulting mispricing \cite{OracleInDece}.

The main defense methods include the M-of-N reporter mechanism, limiting the price change range, time-weighted average price, increasing the minimum transaction delay, etc.

\begin{itemize}
    \item M-of-N Reporter Mechanism
Using multiple oracle providers and adopting the method of calculating the median price is a common way to reduce oracle risks. By obtaining prices and other off-chain data from different oracle providers such as Chainlink, Coinbase, etc. and using the median of these data to calculate the final result, the risk that any single oracle provider's error or malicious operation affects the final result can be effectively avoided. At the same time, to further improve the security of data verification, the project team can also set a threshold to ignore oracle quotes that deviate too much. For example, if the quote of the FTX centralized oracle exceeds the median price by more than 30 basis points, this quote can be regarded as an outlier and ignored. This can reduce the possibility of being attacked and improve the overall data quality and credibility. It should be noted that using multiple oracles also has some potential problems. For example, if there is a collusive attack or conspiracy behavior among multiple oracles, they may provide the same incorrect result to the Dapp, thus attacking the entire system. 

\item Limiting the Price Change Range
By limiting the price to fluctuate within a certain range, the risk of attackers conducting arbitrage by manipulating prices can be effectively avoided. However, if the price changes significantly for some reason while the oracle quote does not change, serious market distortions may occur. For example, during a specific period, the price of an asset suddenly rises or falls a lot, but due to the limitation of the oracle, it cannot reflect this price change in time. This may threaten the solvency of the entire system and then damage the interests and trust of users.

\item Time-Weighted Average Price (TWAP)
Uniswap v2 maintains cumulative prices weighted by elapsed time. A caller can estimate a TWAP by comparing two accumulator observations and dividing by the elapsed interval. Longer averaging windows increase the cost of manipulation but yield less current prices.\cite{UNISWAP} This tradeoff makes TWAP a mitigation whose effectiveness depends on the application and observation interval, rather than a universal prevention of flash-loan attacks.

\item Increasing the Minimum Transaction Delay
Oracle attack arbitrage is a time-sensitive operation, as arbitrageurs typically monitor the market for opportunities to exploit inefficiencies. To minimize risks, attackers often aim to complete both transactions needed to manipulate the oracle price in a single operation, preventing arbitrageurs from intervening in the process. Protocol developers can mitigate this type of oracle attack by introducing a minimum delay. By adding a waiting period between a user's entry and exit from the system, during which the user is temporarily prohibited from trading, the protocol ensures that even if an attacker successfully manipulates the oracle's results, they won't have enough time to perform arbitrage, thereby reducing the impact of the attack on smart contracts.

\end{itemize}

\subsection{Off-chain Factors}

\subsection{Key Leakage}

Private-key compromise can arise from insufficient entropy in key generation. In its disclosure concerning Profanity, 1inch described a 32-bit seed space used to generate 256-bit private keys and demonstrated that keys could be recovered from generated vanity addresses.\cite{ProfanityDisclosure} This weakness concerns effective seed entropy, rather than the visible number of characters in an address. Cryptographically secure key generation and appropriate key-storage controls are therefore necessary to reduce exposure.

\subsection{Hacking of Promotion Software}

Smart contract developers need to communicate and promote their projects to users, and promote projects and interact with users through various social media platforms Like Discord, Twitter, etc.. However, if these social media platforms are hacked or affected by other security issues, it may lead to hackers publishing incorrect information, thereby deceiving users into conducting arbitrage operations and even causing more serious security problems.

\section{Principles of Security Analysis Tools}

Smart contract security analysis plays a crucial role in various mainstream DApps, helping to avoid vulnerabilities in smart contracts. However, as the scale of DApps continues to expand and the complexity of their internal rules increases, traditional manual audit methods can no longer meet the requirements, with both efficiency and vulnerability detection rates declining. To improve the efficiency and accuracy of security analysis, in recent years, smart contract security analysis tools have become a new research direction. Solana smart contract security analysis relies on various security analysis tools, which analyze Solana smart contracts from different angles through static analysis, dynamic analysis, symbolic execution, and other aspects of contract analysis to identify vulnerabilities.

This chapter will briefly introduce the types, principles, and detection methods of smart contract security analysis, with a focus on analyzing important content such as static analysis, dynamic analysis, and symbolic execution. By studying various key technologies, analyzing the advantages, disadvantages, and limitations in practical applications, finding the starting point to improve the ability and efficiency of security analysis, exploring the impact of different methods on security analysis, a full understanding of the principles of different tools can help us further enhance the security of Solana smart contracts.

\subsection{Static Analysis}

Static analysis is one of the earliest automated software security analysis technologies. It refers to the technology of performing security analysis on a program's source code, binary code, etc., without running the software and code. Common methods in static analysis include lexical analysis, data flow analysis, model checking, and theorem proving. Lexical analysis involves scanning the program code to find matching content. If code identical to that in the vulnerability database appears, an alert is issued. Data flow analysis, also known as control flow analysis, uses Abstract Syntax Tree (AST) technology to abstract the program into syntax trees or control flow graphs, enabling rapid analysis of information in the code. Model checking involves constructing a program state model and recording the state transitions of the code during this process to verify whether the code meets certain specific models and characteristics. Theorem proving refers to the technology of transforming the logic in the code into corresponding mathematical formulas and performing solutions and proofs.

Among them, model checking and theorem proving have relatively high accuracy, but they require a large amount of preliminary preparation and the accumulation of a model library. When detecting large programs, their efficiency is relatively low.

Program static analysis refers to generating an intermediate representation of the source code through compilation technology and then analyzing it using formal methods. However, this method has low efficiency. Therefore, researchers often use sharding techniques. In more specific scenarios, such as Crypto API abuse detection, a DAG graph is constructed to analyze the project from front to back. At the same time, the search depth is limited, and the algorithm is shard-processed to reduce the overall complexity. To improve the accuracy of static analysis and reduce the false positive and false negative rates, researchers have been working hard to improve static analysis technology.

Securify provides another static-analysis approach: it derives semantic facts from bytecode and checks compliance and violation patterns.\cite{Securify} This differs from exploring paths by symbolic execution and should be classified accordingly.

Some well-known smart contract static analysis tools include Slither\cite{Slither}, Kudelski Semgrep et.al. Slither is a static smart contract analysis tool suitable for Ethereum. Its working method is as follows:
1. The input of Slither is the smart contract source code, which becomes the corresponding bytecode after being compiled by solc. Slither then obtains the AST through syntax analysis.
2. After information extraction, Slither generates the contract's inheritance graph, control flow graph (CFG), and function list.
3. By converting the contract code into the internal representation language SlithIR, Slither can achieve high-precision analysis, support taint and value tracking, and thus detect complex models.
4. During the code analysis phase, Slither runs a set of predefined analyses, including the dependency relationships between variables and functions in the contract, the read and write operations of variables, and the permission control of functions.
5. Finally, Slither provides functions such as vulnerability detection, code optimization detection, and code understanding output.

\subsection{Dynamic Analysis}

Dynamic analysis, as the name implies, is a technology that analyzes code while the program is running. Security personnel often run the code and modify parameters or intermediate parameters to make the program encounter abnormal situations. Once an abnormal situation occurs, the error information, input, and the on-site environment at that time will leave certain information, enabling researchers to obtain the vulnerabilities and deficiencies in the program from them. Popular dynamic testing techniques include fuzz testing and dynamic taint analysis .

In recent years, researchers have proposed dynamic taint analysis, which tracks polluted data during runtime and marks other variables that interact with it. After the program is completed, the corresponding variables are checked to detect abnormal calls. The principle of this technology mainly includes three aspects: dynamic taint marking, dynamic taint tracking, and illegal operation checking.

Dynamic taint tracking technology still has many deficiencies. First, it cannot handle defects that only modify data. Second, it cannot achieve coverage of most paths and can only analyze one path at a time. Finally, even if an abnormal situation occurs, it is troublesome to find the corresponding code vulnerability.

Compared with dynamic taint tracking technology, fuzz testing has much lower operating and learning costs. Security personnel only need to find the parameters input into the program and then specify the format and range of the parameters. Then the fuzz testing program runs, continuously generating random parameters that meet the requirements and running the program repeatedly. When the program encounters a panic, we can analyze the corresponding exception to find the corresponding vulnerability.

\begin{figure}[h]
    \centering
    \includegraphics[width=0.8\textwidth]{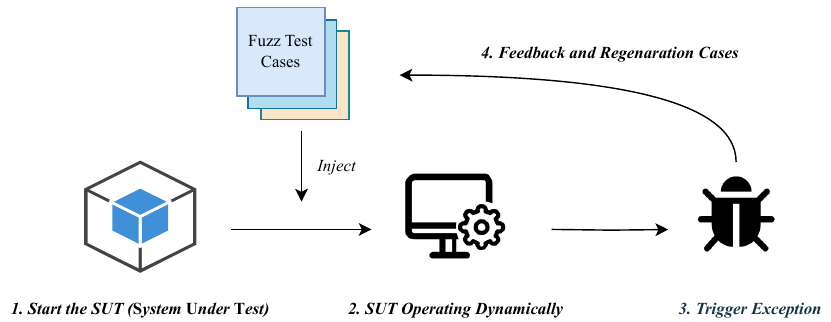}
    \caption{The Procedure of Fuzz Test.}
    \label{fig:FuzzTest}
\end{figure}

As shown in the fig .\ref{fig:FuzzTest}, there are five steps in fuzz testing:
Provide random parameters: Initial parameters are randomly generated and must have the same format and structure as the expected input data of the target program.
Program execution: The provided initial input is fed into the target program, allowing the program to execute.
Exception detection: Monitor and detect exceptions generated during program execution, such as program crashes and assertion failures.
Input mutation: Mutate the successfully executed input to generate new inputs. These mutations can include adding or deleting data, changing data types and formats, etc.
Record exceptions: Record all inputs that cause program exceptions for subsequent analysis and debugging.

By continuously iterating through the above steps, fuzz testing can uncover various defect vulnerabilities in software, thereby helping developers improve software quality.

For Rust components, \texttt{cargo-fuzz} provides integration with libFuzzer.\cite{CargoFuzz} Solana-aware testing additionally requires an execution harness and account setup, as provided by Trident.\cite{Trident} For Ethereum contracts, Echidna generates transaction sequences to test user-specified properties.\cite{Echidna} These approaches illustrate the distinction between finding a runtime failure and checking an application-specific invariant.

\subsection{ Symbolic Execution}
Symbolic execution represents inputs as symbolic values and accumulates constraints along execution paths. Solving these constraints can produce inputs that exercise selected behaviors, while path explosion and environment modeling limit scalability. Oyente applies symbolic execution to Ethereum smart contracts, and Manticore supports symbolic analysis of binaries and Ethereum contracts.\cite{Oyente,Manticore} Solana analysis requires corresponding models for accounts, instructions, and cross-program calls.

Dynamic symbolic execution tools execute the target program and construct path predicates by symbolically interpreting program instructions. Path predicates contain branch conditions encountered during the analysis. The symbolic engine attempts to invert each branch in the path predicate to discover new execution paths that are difficult to reach through fuzz testing. The predicate used for branch inversion connects all previous branch constraints (i.e., the constraints of branches executed before the target branch) and the negation of the target branch constraint. Most symbolic engines often encounter over-constraint and under-constraint problems (similar to taint analysis), which prevent them from exploring more program paths. Over-constraint means that there are many redundant constraints in the path predicate, which may complicate it or even make it unsatisfiable. Over-constraints increase the number of symbolic instructions during the analysis. Conversely, under-constraint means that some variables are not considered symbolic, even though they should be. Non-trivial branch conditions or symbolic pointers (dependent on user data) may lead to under-constraint. Therefore, dynamic symbolic analysis still has many difficulties to overcome.

Oyente explores possible execution paths to identify vulnerability patterns in Ethereum contracts.\cite{Oyente} Securify, discussed under static analysis, instead reasons over semantic facts and compliance or violation patterns.\cite{Securify} These different mechanisms should be distinguished when comparing tool coverage and limitations.

\subsubsection{Combination of Technologies}

In software vulnerability mining techniques, each technique has its own advantages and disadvantages and cannot cover all program paths, especially when dealing with large software projects. Therefore, combining complementary techniques can improve the exploration of relevant behaviors. SMARTIAN combines static transaction-sequence prediction with dynamic data-flow feedback to guide fuzzing.\cite{SMARTIAN} COBRA performs interaction-aware bytecode-level vulnerability detection, highlighting the importance of modeling contract interactions.\cite{COBRA} By combining the advantages of different techniques, the efficiency and accuracy of vulnerability detection can be improved. For example, in static analysis, symbolic execution can be used to reduce false positives, and taint analysis can be used to handle dynamic inputs. In dynamic analysis, fuzz testing can be used to generate a large amount of input data, and then valid input data can be filtered according to the results of symbolic execution.

Mythril is a symbolic-execution-based analysis tool for EVM bytecode.\cite{Mythril} Its platform support should be distinguished from native Solana analysis. More broadly, tool combinations need to be evaluated against their intended vulnerability classes, execution models, and analysis budgets; combining methods alone does not establish complete detection coverage.

\section{Discussion and Conclusion}

As security issues related to smart contracts increasingly affect people's confidence in blockchain, the security analysis tools for smart contracts are also bound to receive more attention. As an essential part of blockchain security, these tools are held to higher standards in terms of accuracy, efficiency, ease of use, and more. This chapter summarizes the main work and findings of this paper regarding the security analysis of Solana blockchain smart contracts and provides prospects for further work.

The Contribution in this paper includes the following:
\begin{itemize}

\item Collecting and organizing vulnerabilities of Solana smart contracts, analyzing and researching security analysis techniques, and summarizing the advantages and disadvantages of various types.

\item Studying mainstream security analysis tools for Solana smart contracts, conducting detailed comparisons, and evaluating the overall security level of Solana in comparison with Solidity.

\item Specializing in research on the types and tools of Solana blockchain smart contract security analysis technology and comparing its security analysis technology level with Solidity.

\end{itemize}

The paper offers a comprehensive summary of Solana smart contract security analysis but emphasizes the need for higher accuracy and efficiency in vulnerability detection. Further in-depth research is essential in several areas to enhance Solana's smart contract security. Key focus areas for future research include the integration of large language models (LLMs), improving tool usability, adapting security analysis methods, and fostering the development of the Solana security ecosystem.\cite{SurveyTool}

LLMs offer complementary assistance for contract understanding and vulnerability analysis. SCLA augments prompts with control-flow information and semantic facts to improve smart-contract summarization.\cite{SCLA} Bu et al. investigate fine-tuned LLMs for vulnerability detection in real-world DApps.\cite{EnhanceSCVul} GPTScan combines GPT-based reasoning with program analysis to detect logic vulnerabilities.\cite{GPTScan} These studies motivate integrating language models with structural program information, but their results do not establish equivalent effectiveness on native Solana programs. Such use requires platform-specific evaluation and validation of generated findings.

Additionally, many smart contract security analysis tools currently lack user-friendliness. Most of these tools remain in source code form without a fully configured environment, creating significant barriers for users. Improving usability is crucial, and contributions from open-source developers and professionals are needed. Furthermore, the Solana smart contract security environment still lags behind Solidity, and code migration methods may be a low-cost way to adapt Solidity security analysis techniques for Solana. The development of EVM compatibility, such as Neon on Solana, could also help bridge this gap by enabling developers to use Ethereum's security tools, although Neon’s development is still ongoing.

\bibliographystyle{ACM-Reference-Format-arxiv}
\bibliography{Ref}

\end{document}